%% file: paper_smartspace.tex
\documentclass[twocolumn,12pt]{IEEEtran}
\usepackage{booktabs}
\usepackage[nolist]{acronym}
\usepackage{amsmath,graphicx}
\usepackage{amsmath,amssymb}
\usepackage{amsthm}

\usepackage{amsmath,amssymb}
\usepackage{algpseudocode}
\usepackage{algorithm}
\usepackage{float}
\usepackage{color}
\usepackage{graphicx}
\usepackage{epstopdf}
\usepackage{epsfig}
\usepackage{icomma}
\usepackage{subcaption}
\usepackage{colortbl}
\usepackage{hhline}
\usepackage{algorithmicx}
\usepackage{balance}
\usepackage{multicol}
\usepackage{soul}
\usepackage{xcolor}

\usepackage{mathtools}
\usepackage[top=0.90in, bottom=1.2in, left=0.673in, right=0.667in]{geometry}

\usepackage{cite}

\ifCLASSINFOpdf
\else
\fi
\begin{document}

%
\title{ML-based PBCH symbol detection and equalization for 5G Non-Terrestrial Networks}
%
%
%

\author{\IEEEauthorblockN{ Inés Larráyoz-Arrigote$^\dagger$, Marcele O. K. Mendon\c{c}a$^\star$, Alejandro~Gonzalez-Garrido$^\star$ , Jevgenij~Krivochiza$^\star$, Sumit~Kumar$^\star$, Jorge Querol$^\star$,  Joel Grotz$^\P$, Stefano Andrenacci$^\P$, Symeon~Chatzinotas$^\star$ } \\
\IEEEauthorblockA{
\normalsize{\textit{$^\star$SnT - University of Luxembourg, Luxembourg.} } }\\ 
\IEEEauthorblockA{
\textit{$^\dagger$Universidad Pública de Navarra (UPNA), Navarra, Spain}, \textit{$^\P$SES S.A., Luxembourg}} \\
\IEEEauthorblockA{
\textit{Corresponding Author:  marcele.kuhfuss@uni.lu.} }

}

\maketitle
\thispagestyle{empty}
\begin{abstract}
This paper delves into the application of Machine Learning (ML) techniques in the realm of 5G Non-Terrestrial Networks (5G-NTN), particularly focusing on symbol detection and equalization for the Physical Broadcast Channel (PBCH). As 5G-NTN gains prominence within the 3GPP ecosystem, ML offers significant potential to enhance wireless communication performance. To investigate these possibilities, we present ML-based models trained with both synthetic and real data from a real 5G over-the-satellite testbed. Our analysis includes examining the performance of these models under various Signal-to-Noise Ratio (SNR) scenarios and evaluating their effectiveness in symbol enhancement and channel equalization tasks. The results highlight the ML performance in controlled settings and their adaptability to real-world challenges, shedding light on the potential benefits of the application of ML in 5G-NTN.
\end{abstract}


%
\IEEEpeerreviewmaketitle

\setlength\doublerulesep{0.7pt}  

\section{Introduction and Background}
\label{sec:intro}
The application of \ac{ML} techniques in wireless communications is continuously proving its enormous potential towards performance enhancement and acceleration of complex signal processing algorithms. Recently \ac{NTN} has gained significant momentum among the research community, especially after the inclusion of \ac{NTN} as a part of 3GPP ecosystem from the recent Release-17 on-wards\cite{5gamericas}. Besides, 3GPP Release-18 \cite{lin2022overview} will natively embrace artificial intelligence and machine learning based technologies for providing data-driven and, intelligent network solutions. 

Several studies have explored the application of ML in the context of 5G-NTN.
The authors in \cite{machumilane2023learning} apply reinforcement learning to determine appropriate scheduling policy for link selection in a LEO based 5G-NTN. Their simulations show the effectiveness of this approach in terms of improving end-to-end loss rates and bandwidth utilization for a non-static channel. Authors in \cite{dahouda2023machine} focus on application of ML techniques to address the problem of handovers in a LEO based 5G-NTN. Location of the UE is taken as the important feature to train the ML model for improving the conditional handover decisions.  In a survey article \cite{mahboob2023tutorial}, the authors have provided deep insights into the applications of artificial intelligence (AI) empowered techniques for 5G and 6G NTN which include: channel estimation, mobility management, doppler estimation and compensation, resource management, network procedures to name a few. 
Moreover, ML has been extensively investigated in order to address fundamental physical layer challenges in wireless communication systems.  In \cite{ye2017power}, joint channel estimation and symbol detection is performed by one DNN in an end-to-end manner. However, domain-specific knowledge is exploited in  \cite{gao2018comnet, mendoncca2021ofdm}  by breaking  a single DNN in two. Notably, these studies primarily rely on simulations and lack real-world data validation, highlighting the need for practical testing. 

In this work, we focus on ML for symbol detection and equalization in 5G-NTN's Physical Broadcast Channel (PBCH). The PBCH plays a crucial role in conveying essential data, via the \ac{MIB} which is necessary for initial access procedure in the \ac{UE}; this includes acquisition of System Information Block-1 (SIB1) and the location of resources in Physical Downlink Control Channel (PDCCH). To the best of author's knowledge, ML based techniques has not been used for symbol detection in 5G-NTN over real world data. We benefit from the 5G-NTN testbed at the University of Luxembourg to record live Over-the-Satellite (OTS) IQ samples for the functional validation and performance verification of our ML algorithms.  

\section{System model and Testbed} \label{sec:system}

\subsection{5G NR Synchronization System Block}

This research aim to improve the \ac{MIB} decodification. In \ac{5G}, the \ac{MIB} is embeded into the \ac{SSB}, and to help the \ac{UE} decoding it, it is send in bursts. These burst have a period of 20~ms in most of the configurations, and each burst consist on several repetitions of the \ac{SSB}. This number of repetitions also is controlled by the system configuration. All these parameters are described in detail in the \ac{5G} standard document \cite{3gpp_nr_nodate-8}.

\begin{figure}[!tb] \centering \includegraphics[width=9.0cm]{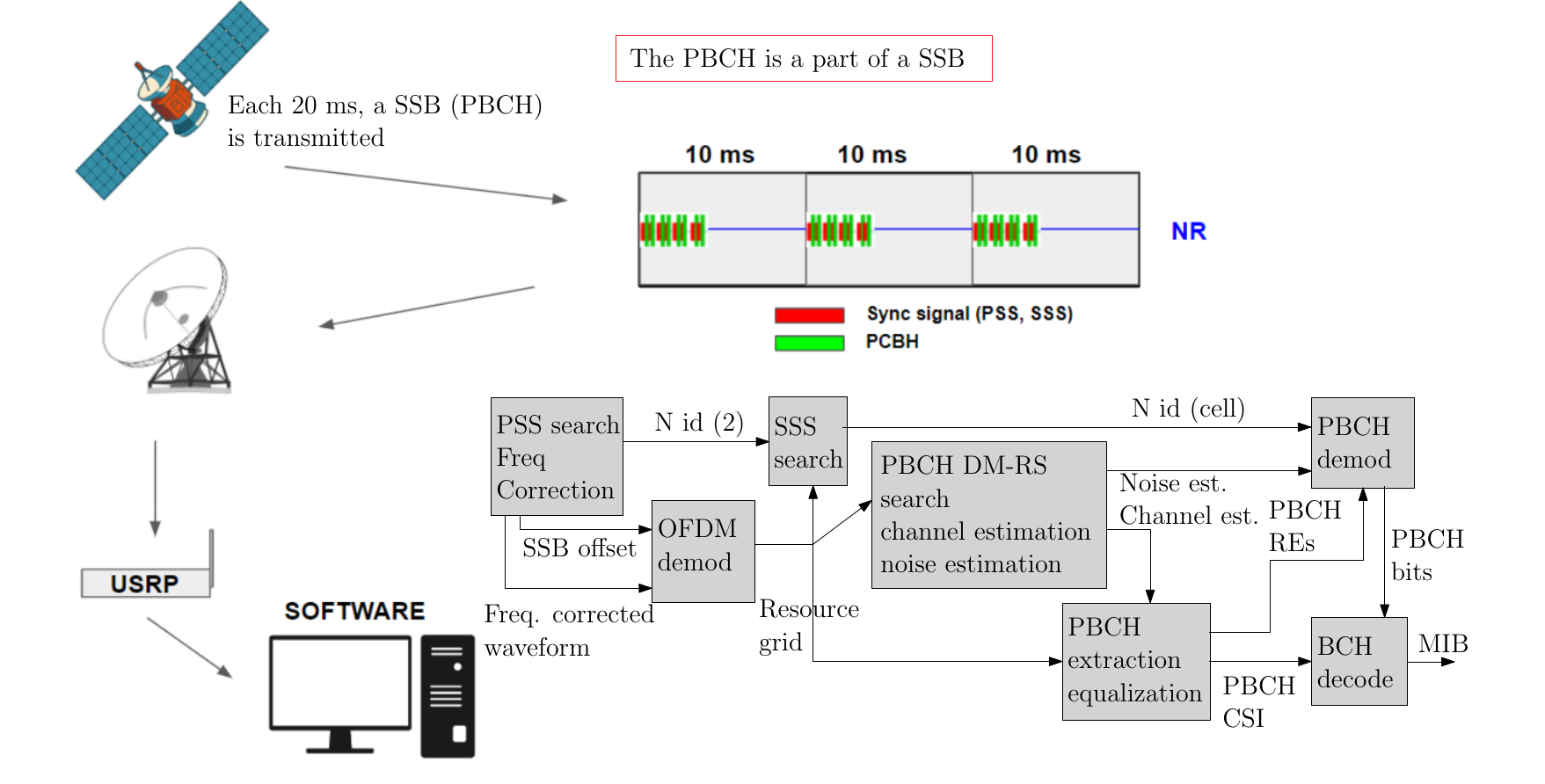} \caption{ Block diagram of the 5G signal capturing and decoding system with proposed NNs. } \label{fig:problem_ilust} \end{figure}

Therefore, the first task of the \ac{UE} is to find in the \ac{RG} the position of the \ac{SSB}. Fig. \ref{fig:problem_ilust} shows the receiver designed to extract the \ac{MIB} from the received signal using a \ac{ML} enhanced equalization. The receiver functionalities are described in the following paragraphs, whereas the \ac{ML} model is further detailed in the next section.

The process to locate the \ac{SSB} in the \ac{5G} signal starts by a blind search of the \ac{PSS} and the \ac{SSS} based in the \ac{GSCN} raster. In our experiment, we skip the blind search using the \ac{GSCN} as we have control of the transmitter. 

Once we have found the \ac{PSS} and \ac{SSS}, the receiver knows the position in time and frequency of the \ac{SSB} within the \ac{RG} and a coarse estimation of the \ac{CFO}. With this information, the receiver is able to locate the \ac{RE} that correspond to the \ac{MIB} and the ones used for the \ac{DMRS}.

The next step in our receiver is to estimate the channel. In \ac{5G} this is done by using the \ac{DMRS} pilots. Our receiver uses these pilots for the \ac{NN} equalization and further enhancement of the \ac{MIB} decoding process. This process is the core of this research and is detailed in the next section.

\subsection{Testbed}


    


We are using a system composed of a USRP (N310 model), which is a software-defined radio device that can capture and transmit various types of wireless signals over a wide range of frequencies. The USRP is connected to a laptop via an Ethernet cable, which allows us to control the USRP settings and process the captured signals using MATLAB. The USRP is used to capture 5G signals from a terrestrial base station located at the 6GSpace Lab, which is a research facility that aims to develop and test innovative solutions for future wireless communications. The base station has an antenna that targets the SES satellites in geostationary orbit, which provide global coverage and high data rates for 5G services.

The captured signals contain SS/PBCH blocks that carry the MIB, which is mandatory system information that provides basic parameters for initial cell selection and access. To decode the MIB, we need to perform several steps using MATLAB.

We configure the USRP to capture a certain length of samples at a specific frequency and gain. The center frequency of the 5G signal is 2029.25 MHz, but there is an offset between the carrier and the USRP, so we set the configured frequency to 2029.2 MHz. This offset introduces some phase error in the received signal, which we need to compensate for in the later stages. We also need to choose a capture length that is long enough to contain at least one SS/PBCH block, which occurs every 20 ms according to the 5G standard.

\section{Methodologhy}\label{sec:met}
\begin{figure}[!tb]
         \centering
        	\includegraphics[width=8.5cm]{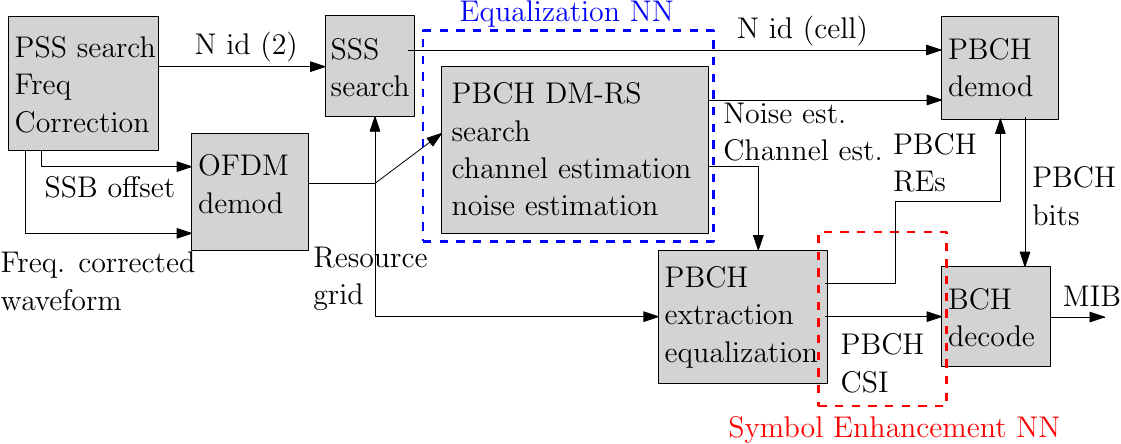}
        \caption{Block diagram of the 5G UE PBCH receiving chain with proposed NNs}
        \label{fig:diag_proposed_NNs}
    \end{figure}

We have considered two distinct \ac{NN} algorithms to enhance various aspects of the on-ground 5G \ac{UE} receiving chain. The first one, denoted as Symbol Enhancement NN, is dedicated to refining received symbols post-equalization. The second one, referred to as Equalization NN, is designed to improve channel estimation and, critically, the equalization process itself. Figure~\ref{fig:diag_proposed_NNs} depicts the block diagram of the 5G UE PBCH receiving chain indicating where the proposed NNs are placed within it.

The Symbol Enhancement NN just focuses on enhancing symbols following the equalization process, performing a task of relatively low complexity. Conversely, the Equalization NN needs to perform a more challenging task due to the multifaceted nature of its objective (i.e. joint channel estimation, noise characterization, equalization and the enhancement of the equalization process applied to received data).

In subsection \ref{subsec_arq}, we describe the model architecture and the training process common to both NNs.  We use synthetic and real data from our 6GSPACE Lab testbed \cite{6GSPACELab} for training, with specific procedures described in subsections \ref{subsec_synt_data} and \ref{subsec_real_data} for synthetic and real data, respectively.

\subsection{NN architecture and training process}\label{subsec_arq}
For both NNs, we employ a fully connected architecture with three hidden layers, each containing $K$ neurons. The hidden layers use the hyperbolic tangent activation function, whereas the output layer employs a linear activation function. The NNs take real-valued block versions of the received complex symbols as input. For training, both NNs employ the Adam optimizer with a learning rate of $\mu=0.001$ to minimize the mean squared error (MSE) between transmitted symbols and predictions. The training process employs mini-batches of 50 samples, requiring 40 epochs to train the NNs.

\subsection{Synthetic Data}\label{subsec_synt_data}

Both transmitted and received synthetic data have been generated using MATLAB in order to train and validate the NNs. To generate the transmitted samples, we meticulously defined specific parameters to replicate the 5G signal used in the real data case. These parameters include selecting the SSB "Case A" block pattern, corresponding to a sub-carrier spacing (SCS) of 15~kHz, and establishing a minimum channel bandwidth of 5~MHz. As a channel model, we have considered Additive White Gaussian Noise (AWGN), a carrier frequency offset (CFO) uniformly distributed within the SCS, and an integer and fractional delay corresponding to a GEO satellite delay. Additionally, we introduced the standard variability in the MIB by changing the cell identification number and the frame sequence number on the synthetically generated 5G samples.

Both NNs were trained using a dataset comprising 3024 transmitted and received SSB signals, with each dataset tailored to a specific SNR. For the Symbol Enhancement NN, the ML model was trained using the post-equalization symbols, where the channel estimation and equalization were performed with a classical Minimum Mean Squared Error (MMSE) algorithm. On the other hand, the Equalization NN was trained with the symbols just after the SSB synchronization algorithm, leaving the NN in charge of the complete channel estimation and equalization tasks.

Furthermore, to gain a more comprehensive understanding of the performance of these ML models, we performed the ML test using three distinct configurations:
\begin{itemize}
    \item Configuration 1: Test multiple ML models, each trained with a different SNR.
    \item Configuration 2: Test a single ML model trained across a range of SNRs.
    \item Configuration 3: Test a single ML model trained with a fixed SNR of 20~dB.
\end{itemize}
The SNRs considered for Options 1 and 2 included 0 dB, 2 dB, 5 dB, 7 dB, 10 dB, 15 dB, and 20 dB.

\subsection{Real Data}\label{subsec_real_data}

In the real data scenario, several differences arise compared to synthetic data. The 5G UE receiver lacks real-time access to transmitted data due to dynamic factors.
 For example, the frame sequence number is not a priori known at the receiver since it is constantly increasing per each transmitted frame. On the other hand, the channel conditions tend to fluctuate over time and introduce dynamic variations into the experiments. Those are coming from both the satellite payload (e.g. non-linearities) and the over-the-air channel effects (e.g. tropospheric fading).

\begin{figure}[!tb]
         \centering
        	\includegraphics[width=5.5cm]{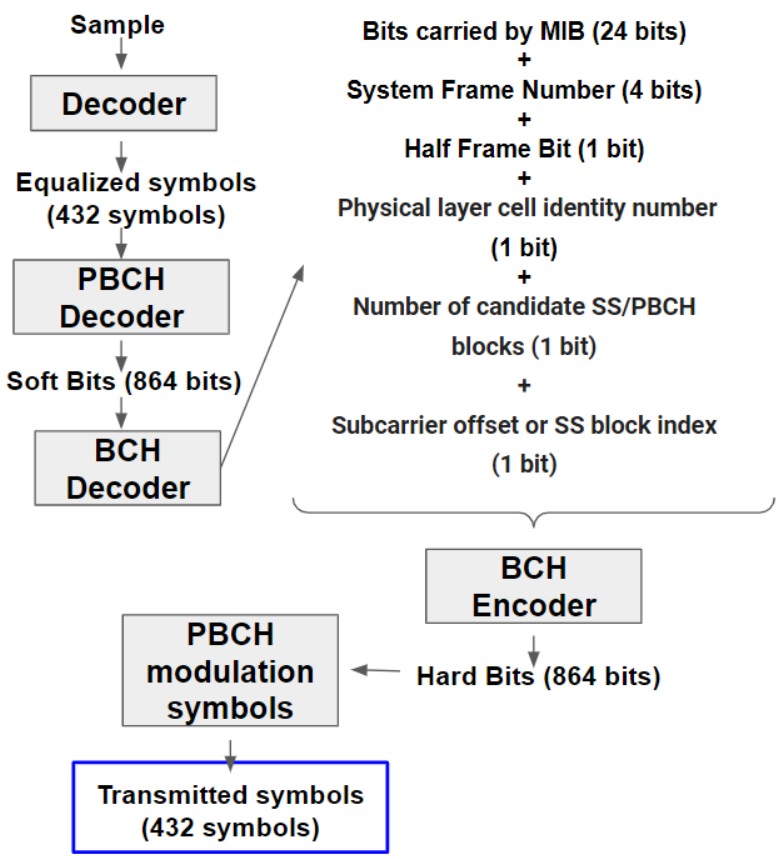}
        \caption{Transmitted real data regeneration-after-decoding approach.}
        \label{fig:tx_data_gen}
    \end{figure}

Perfect knowledge of the transmitted data is required to train the NNs, and as mentioned above, this is not possible with the current setup. To address  this issue, we have adopted the synthetic regeneration-after-decoding approach depicted in Figure~\ref{fig:tx_data_gen}. This process involves decoding the PBCH bits using the 5G standard approach until the 32-bit payload data is obtained. The 32-bit payload data comprises 24 bits corresponding to the MIB and an additional 8 bits from various parameters. In the 5G standard, the 32-bit payload data has attached a 24-bit cyclic redundancy check (CRC) code. If the CRC determines that the 32-bit payload data are correct, the regeneration-after-decoding is executed. The 32-bit payload data are fed into the Bose-Chaudhuri-Hocquenghem (BCH) encoder, yielding 864 bits. These bits are then modulated with a Quadrature Phase-Shift Keying (QPSK) scheme obtaining the 432 symbols originally transmitted through the satellite, which are used for NNs training purposes. The regeneration-after-decoding approach is acceptable in this case since the SNR level at which the decoder operates is larger than the SNR ranges at which the NN are evaluated.

For the testing phase of the NNs with the real data over the satellite, the USRP gain settings were configured to three different values: 70~dB, 25~dB, and 20~dB. Such gain values correspond to SNR values of 20~dB, 10~dB and 3~dB respectively.





\section{Results}\label{sec:results}

In this section, we evaluate the proposed NN models for symbol enhancement and equalization tasks using both synthetic and real data. As both NNs consider the tasks as regression problems, the ML metric used to evaluate the models is the MSE loss.

The NNs produce real-valued complex symbols, which are then converted back into complex symbols. These symbols can be represented as constellations (see Figure \ref{fig:const_synt_data_20_NN1}). We demodulate the complex symbols to obtain received bits, which are compared to transmitted bits to calculate the bit error rate (BER), serving as a system performance metric.

\subsection{Synthetic data}
\begin{figure}[!h]

    \begin{subfigure}[h]{0.24\textwidth}
    \centering
       	\includegraphics[width=4.2cm,trim={0.2cm 0 0 0},clip]{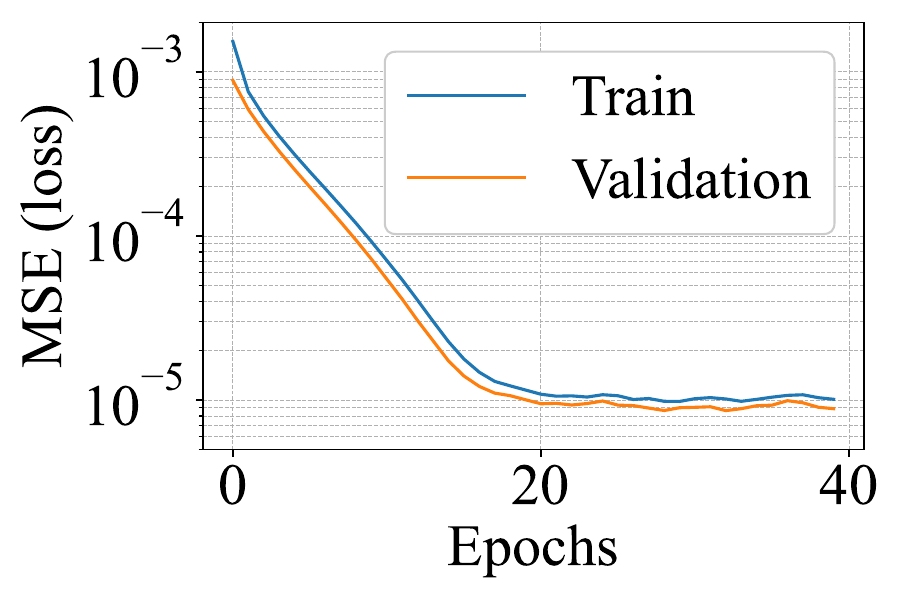}
        \caption{Symbol Enhancement NN }
        \label{fig:lc_synt_data_20_NN1}
    \end{subfigure}
    \begin{subfigure}[h]{0.24\textwidth}
        \centering
        	\includegraphics[width=4.2cm,trim={ 0.7cm 0.8cm 0 0},clip]{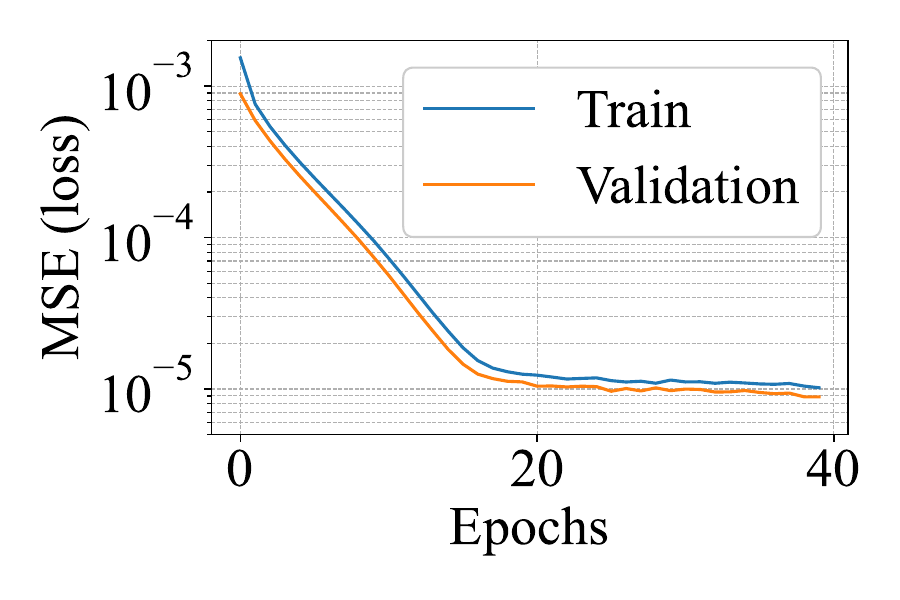}
        \caption{Equalization NN }
        \label{fig:lc_synt_data_20_NN2}
    \end{subfigure} 
      \caption{Learning curves for NNs trained and tested with SNR = 20dB with synthetic data.}
            \label{fig:lc_synt_data_20}
    \end{figure}

\begin{figure}[!h]

    \begin{subfigure}[h]{0.24\textwidth}
    \centering
       	\includegraphics[width=4.2cm,trim={0.2cm 0 0 0},clip]{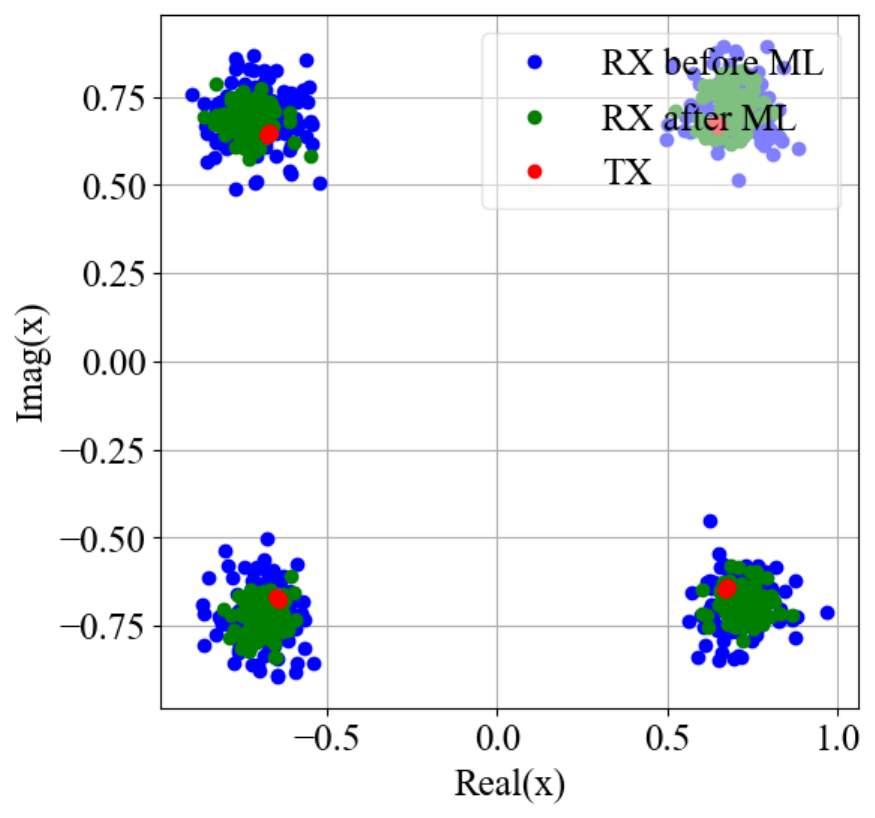}
        \caption{Symbol Enhancement NN }
        \label{fig:const_synt_data_20_NN1}
    \end{subfigure}
    \begin{subfigure}[h]{0.24\textwidth}
        \centering
        	\includegraphics[width=4.2cm,trim={ 0.2cm 0 0 0},clip]{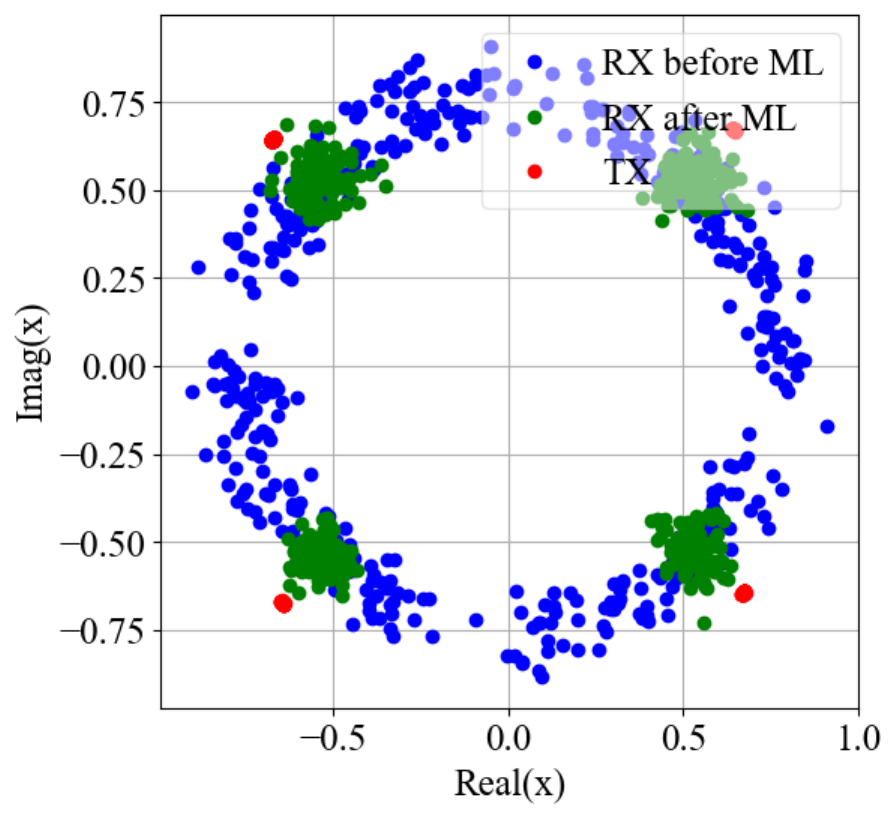}
        \caption{Equalization NN}
        \label{fig:const_synt_data_20_NN2}
    \end{subfigure} 
      \caption{Constellations for NNs trained and tested with SNR = 20dB with synthetic data.}
            \label{fig:const_synt_data_20_2}
    \end{figure}

In this subsection, we consider a synthetic dataset for both the training and testing phases of the proposed NN models. This synthetic dataset is essential for assessing how well our models function in a controlled environment.
The obtained MSE is shown in Figures \ref{fig:lc_synt_data_20_NN1} and \ref{fig:lc_synt_data_20_NN2} for Symbol Enhancement NN and Equalization NN, respectively, trained and tested with SNR = 20 dB. The NNs exhibit optimal performance when evaluated under SNR conditions matching their training data. For instance, improved symbol constellations are evident in Figures \ref{fig:const_synt_data_20_NN1} and \ref{fig:const_synt_data_20_NN2} when the NNs are tested with samples that have the same SNR as the training data.

\begin{figure}[!h]

    \begin{subfigure}[h]{0.24\textwidth}
    \centering
       	\includegraphics[width=4.2cm,trim={0.2cm 0 0 0},clip]{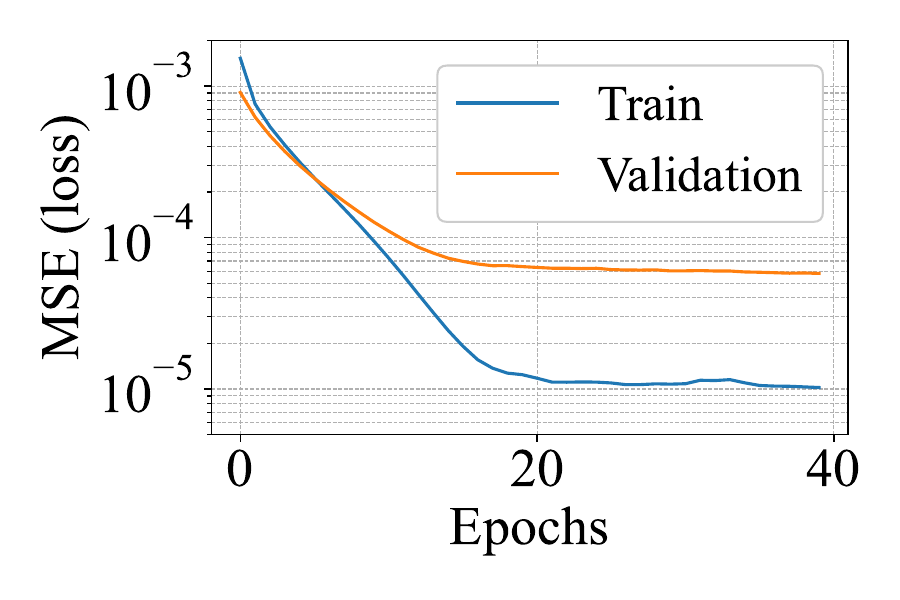}
        \caption{20,10}
        \label{fig:lc_synt_data_NN2_20_10}
    \end{subfigure}
    \begin{subfigure}[h]{0.24\textwidth}
        \centering
        	\includegraphics[width=4.2cm,trim={ 0.7cm 0.8cm 0 0},clip]{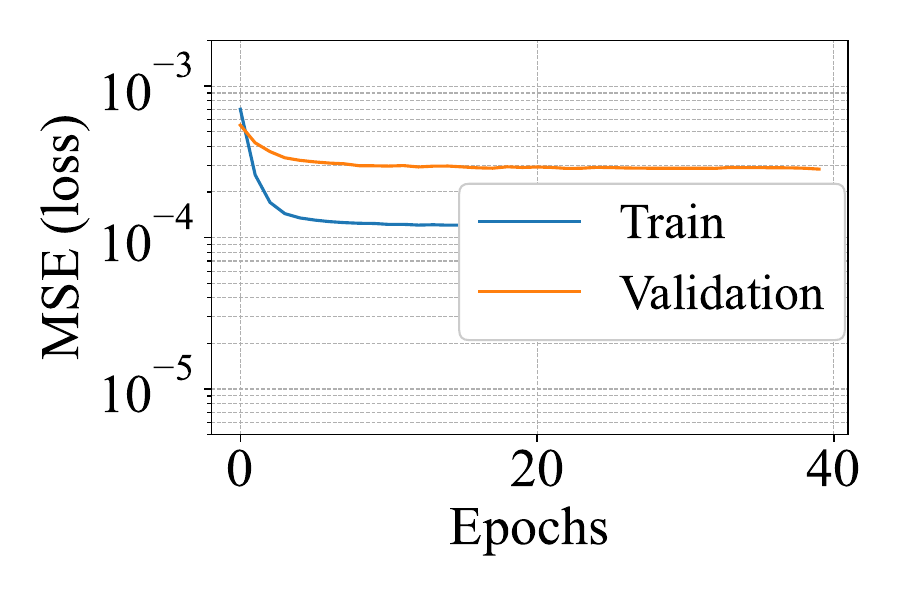}
        \caption{several,20}
        \label{fig:lc_synt_data_NN2_all_20}
    \end{subfigure} 
      \caption{Learning curves for Equalization NN with synthetic data. (a) Trained with SNR = 20 dB and tested with SNR = 10 dB. (b) Trained with several SNRs and tested with SNR = 20 dB}
            \label{fig:lc_synt_data_20_NN2_tests}
    \end{figure}

On the other hand, testing the models at SNRs different from their training data leads to significant performance degradation For instance,
when Equalization NN is trained with samples with a SNR of 20 dB and then tested with samples with a SNR of 10 dB, the  gap between the MSE obtained during training and validation increases, as illustrated in Figure \ref{fig:lc_synt_data_NN2_20_10}.

    A similar degradation in performance can be noticed in Figure \ref{fig:lc_synt_data_NN2_all_20} when the models are trained across a range of SNRs but are evaluated under a specific, fixed SNR condition. 
    

The models' ability to generalize and maintain their efficacy across a variety of real-world scenarios is directly impacted by the nature of the samples used during training.



\begin{figure}[!h]
         \centering
        	\includegraphics[width=8.8cm]{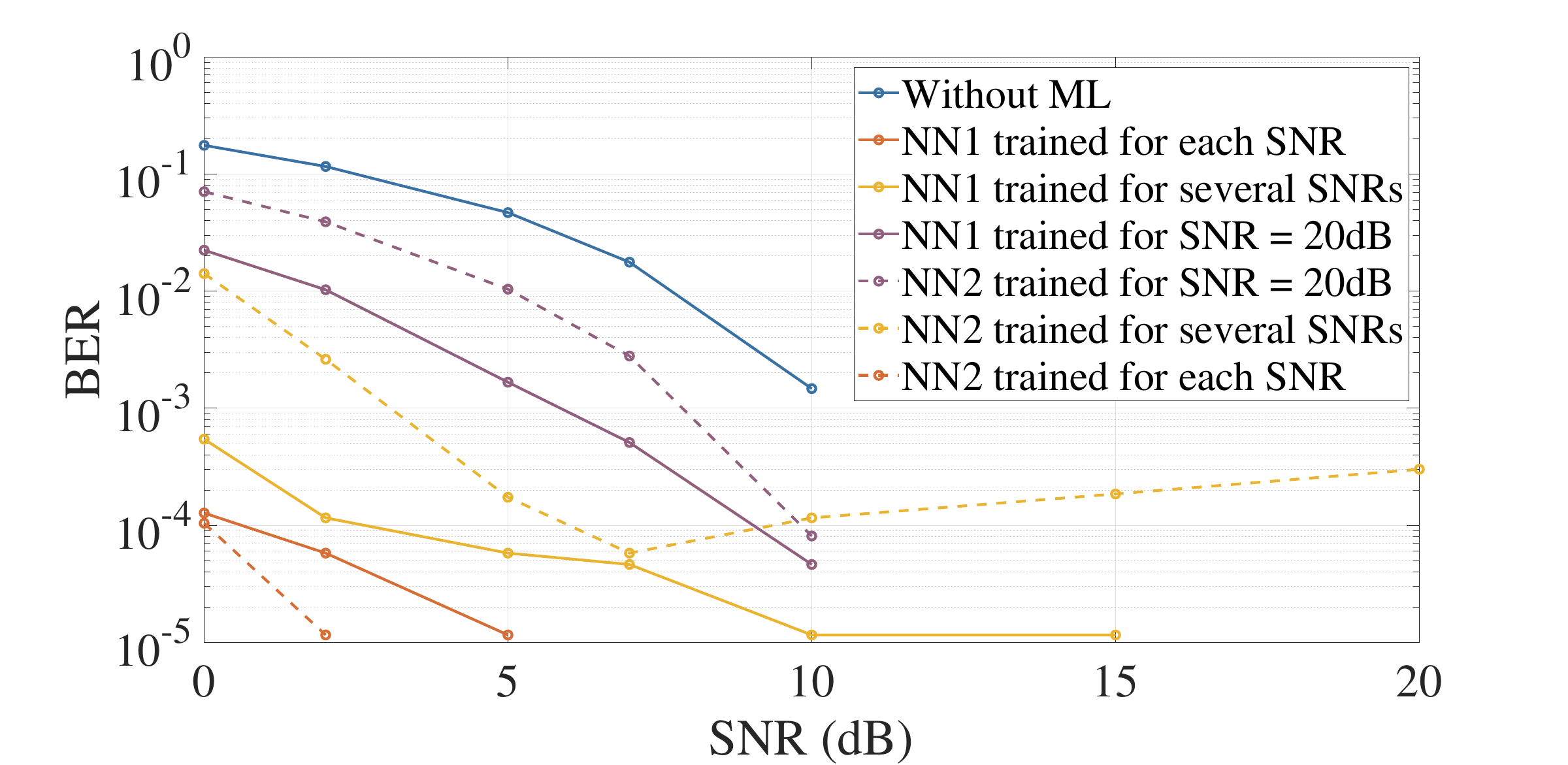}
        \caption{BER curves before and after Symbol Enhancement NN and Equalization NN for synthetic data. }
        \label{fig:ber_synt_data_NN1}
    \end{figure}

We compare the BER without ML techniques to the BER with our NN models in Figure \ref{fig:ber_synt_data_NN1}. We explore three training scenarios. In the first scenario, individual training for each SNR results in specialized models for each SNR setting. In the second scenario, training across various SNRs leads to a single unified model for symbol enhancement and another for equalization, testing the models' adaptability. In the third and final scenario, exclusive training at SNR = 20 dB produces a model specialized for this SNR setting. These trained models are then tested at specific SNR levels.

\subsection{Real data}

\begin{figure}[!h]

    \begin{subfigure}[h]{0.24\textwidth}
    \centering
       	\includegraphics[width=4.2cm,trim={0.2cm 0 0 0},clip]{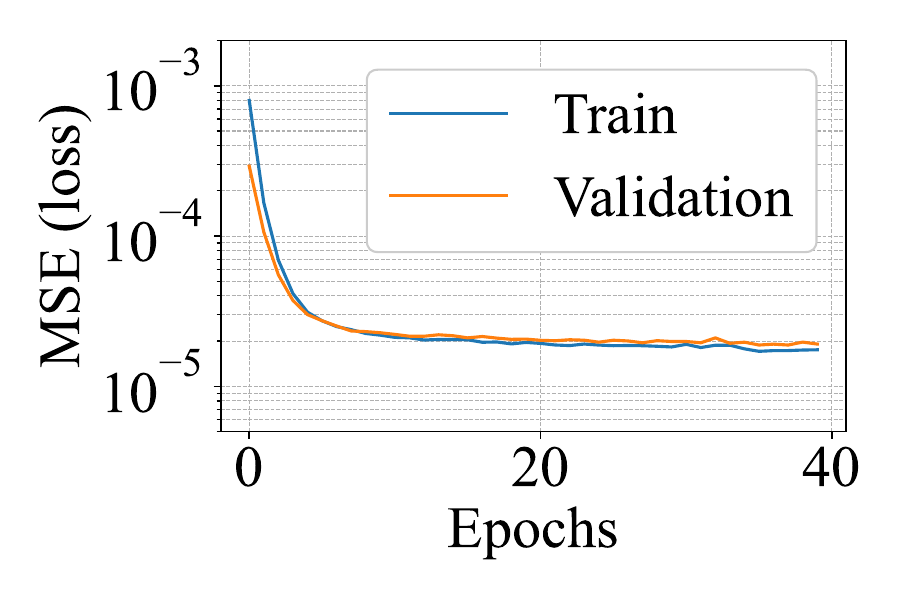}
        \caption{Symbol Enhancement NN }
        \label{fig:lc_real_data_NN1}
    \end{subfigure}
    \begin{subfigure}[h]{0.24\textwidth}
        \centering
        	\includegraphics[width=4.2cm,trim={ 0.7cm 0.8cm 0 0},clip]{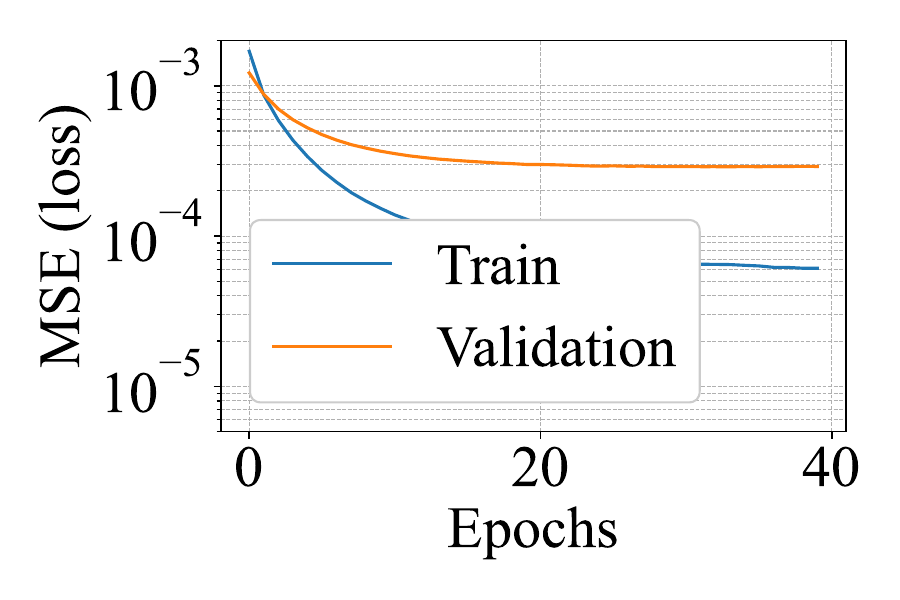}
        \caption{Equalization NN}
        \label{fig:lc_real_data_NN2}
    \end{subfigure} 
      \caption{Learning curves for NNs trained and tested with SNR = 20dB with real data.}
            \label{fig:lc_real_data}
    \end{figure}

In this subsection, we consider a dataset obtained from real-world experimental tests to train and test the proposed NN models. The obtained MSE is shown in Figures \ref{fig:lc_real_data_NN1} and \ref{fig:lc_real_data_NN2} for Symbol Enhancement NN and Equalization NN, respectively. 

\begin{figure}[!h]

    \begin{subfigure}[h]{0.24\textwidth}
    \centering
       	\includegraphics[width=4.2cm,trim={0.2cm 0 0 0},clip]{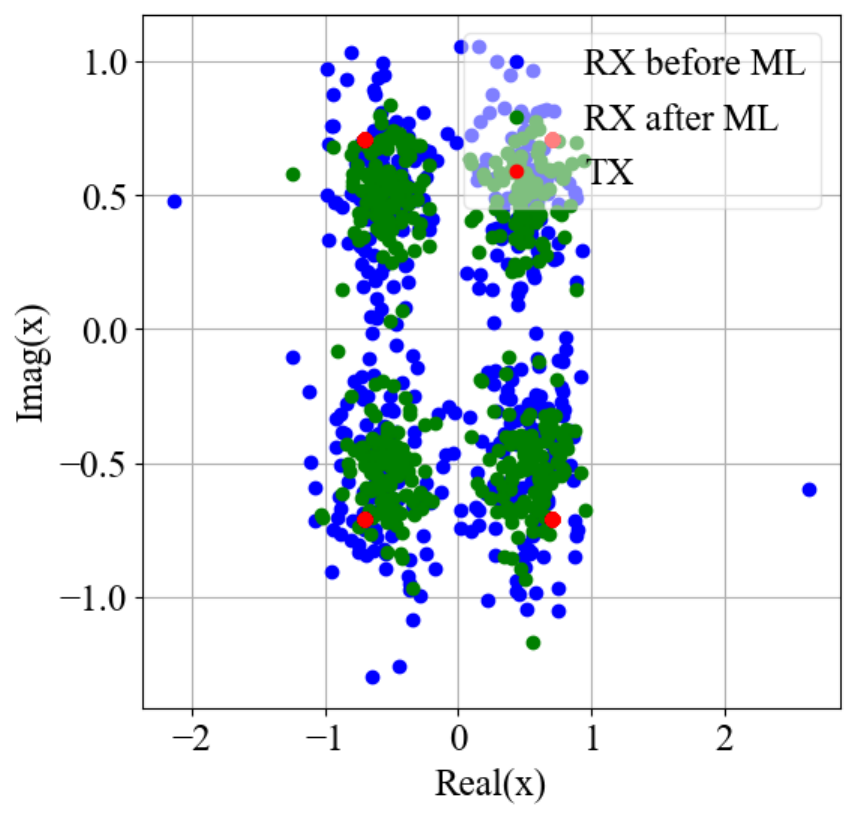}
        \caption{Symbol Enhancement NN }
        \label{fig:const_10_real_data_NN1}
    \end{subfigure}
    \begin{subfigure}[h]{0.24\textwidth}
        \centering
        	\includegraphics[width=4.2cm,trim={ 0.2cm 0 0 0},clip]{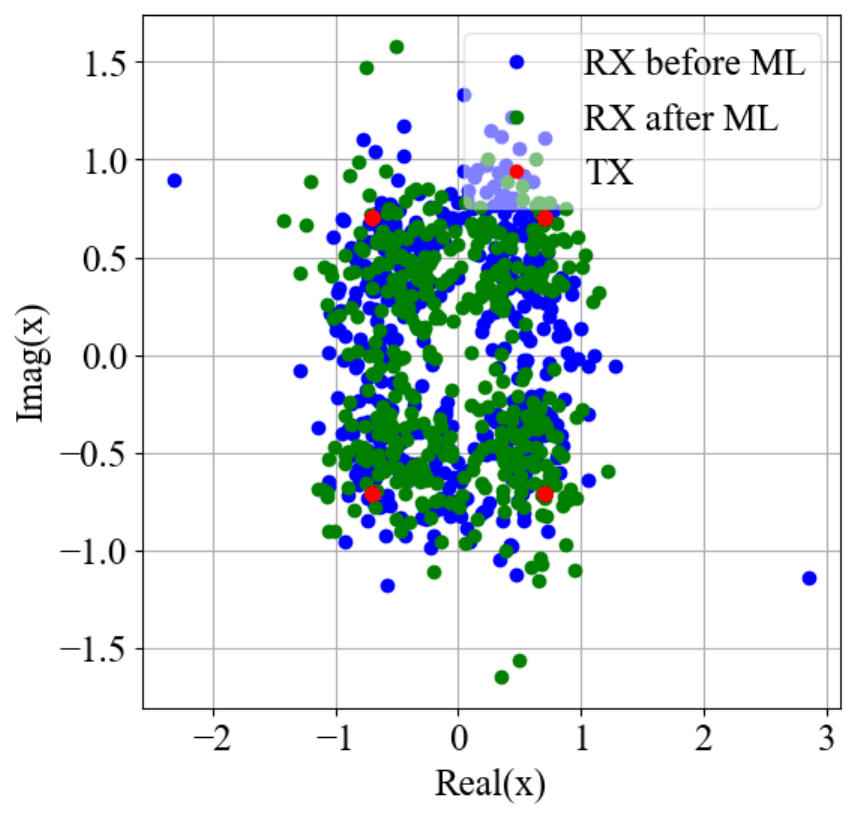}
        \caption{Equalization NN }
        \label{fig:const_10_real_data_NN2}
    \end{subfigure} 
      \caption{Constellation for NNs trained and tested with SNR = 10dB for real data.}
            \label{fig:const_real_data_10}
    \end{figure}

\begin{figure}[!h]
         \centering
        	\includegraphics[width=8.8cm]{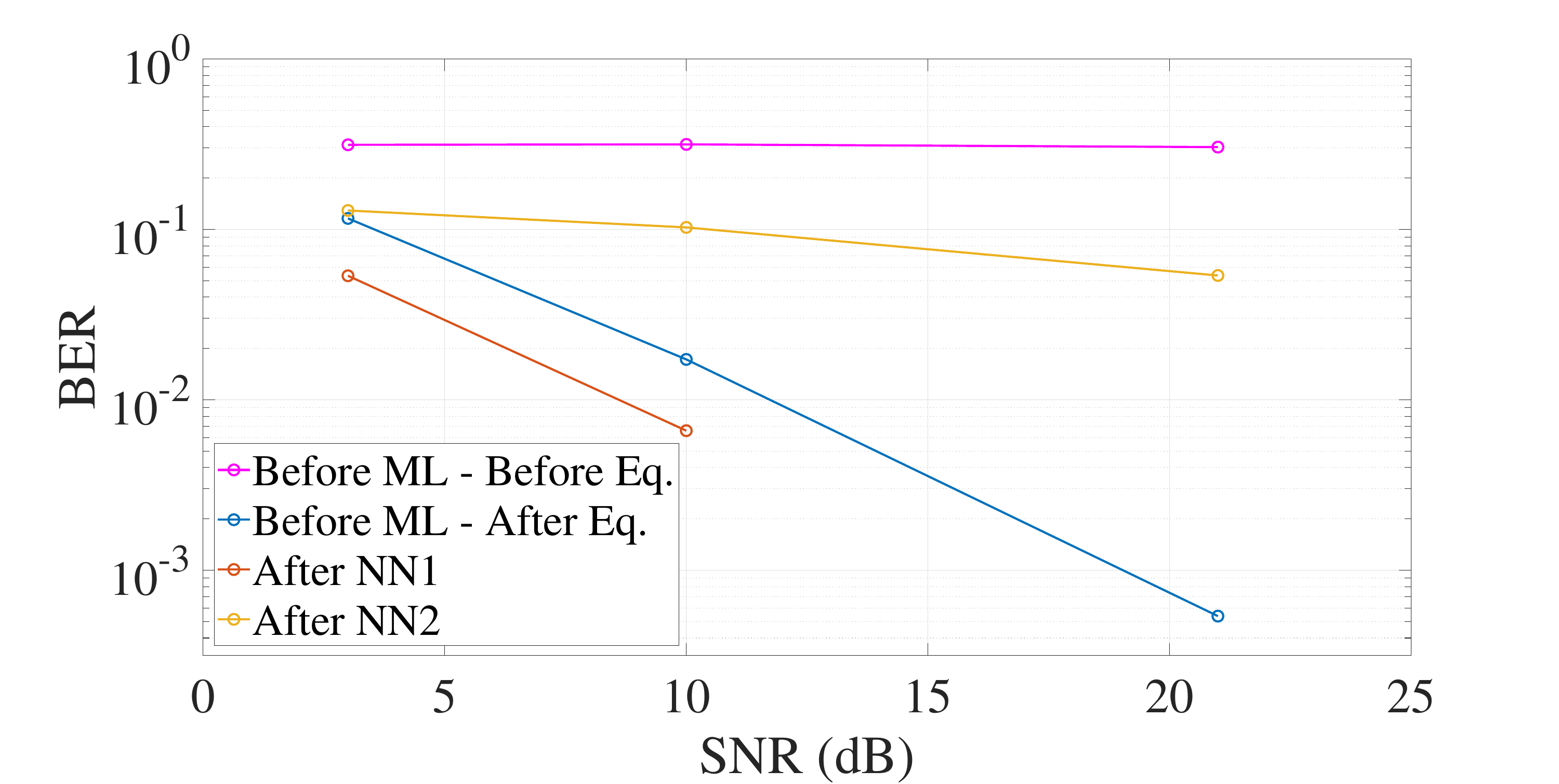}
        \caption{BER curve for real data before and after Symbol Enhancement NN and Equalization NN.}
        \label{fig:ber_real_data_NN1}
    \end{figure}

Symbol Enhancement NN demonstrates a satisfactory level of performance, maintaining acceptable MSE values, and its constellations in Figures \ref{fig:const_10_real_data_NN1} and \ref{fig:const_10_real_data_NN1} closely resemble the transmitted symbols. In terms of BER, Figure \ref{fig:ber_real_data_NN1} shows Symbol Enhancement NN outperforming traditional equalizers without ML. This indicates Symbol Enhancement NN's effectiveness and robustness in real-world data scenarios.

\begin{figure}[!h]

    \begin{subfigure}[h]{0.24\textwidth}
    \centering
       	\includegraphics[width=4.2cm,trim={0.2cm 0 0 0},clip]{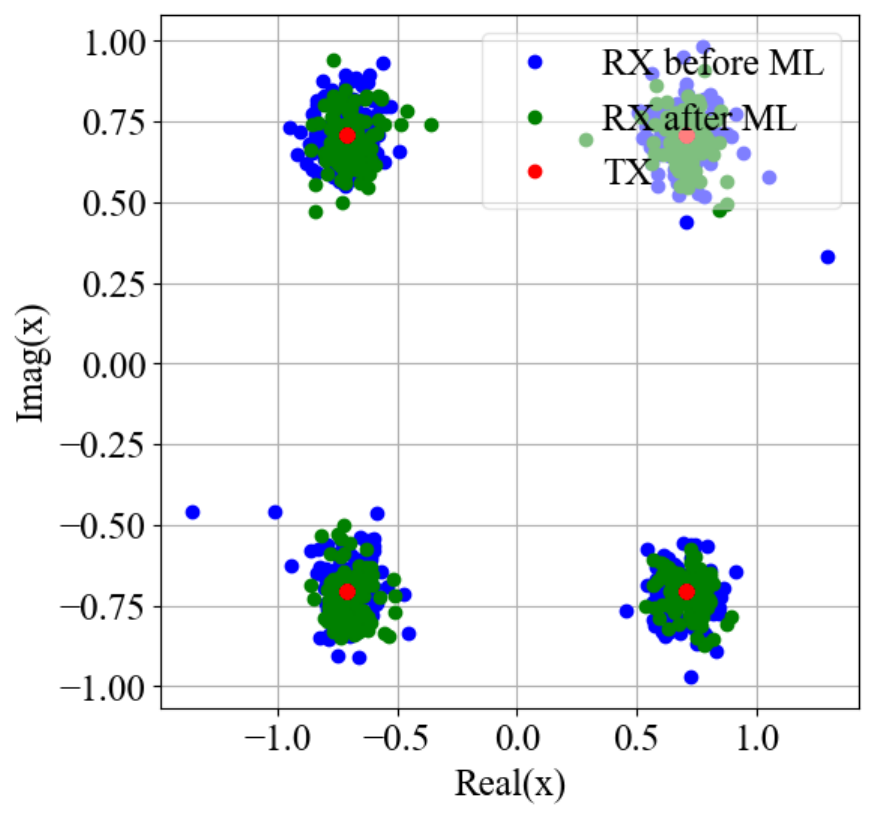}
        \caption{Symbol Enhancement NN }
        \label{fig:const_20_real_data_NN1}
    \end{subfigure}
    \begin{subfigure}[h]{0.24\textwidth}
        \centering
        	\includegraphics[width=4.2cm,trim={ 0.2cm 0 0 0},clip]{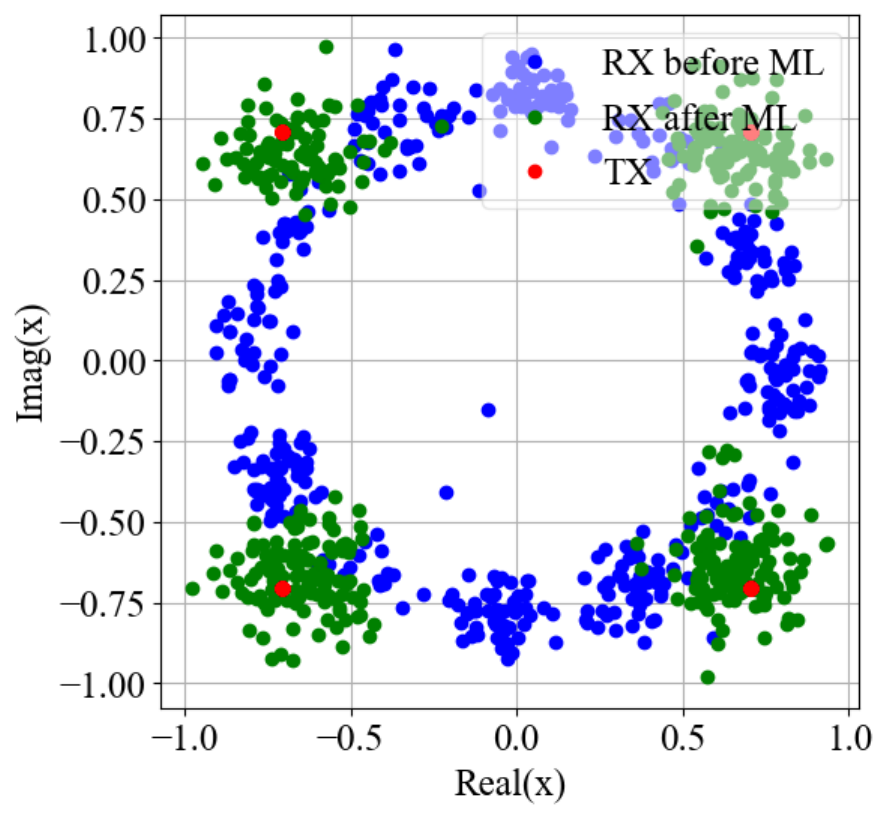}
        \caption{Equalization NN }
        \label{fig:const_20_real_data_NN2}
    \end{subfigure} 
      \caption{Constellations for NNs trained and tested with SNR = 20dB for real data.}
            \label{fig:const_real_data_20}
    \end{figure}

On the other hand, Equalization NN exhibits a larger gap between training and validation MSE in Figure \ref{fig:lc_real_data_NN2}, indicating poorer performance with real data compared to synthetic data. While Figures \ref{fig:const_20_real_data_NN2} and \ref{fig:const_10_real_data_NN2} suggest Equalization NN's ability to equalize symbols, its BER is less favorable than conventional equalizers without ML, as shown in Figure \ref{fig:ber_real_data_NN1}. This underscores the challenge of addressing equalization problems when training data inadequately represents the issues targeted by the Equalization NN.

\section{Conclusion and Future Work}\label{sec:concc}

In this work, we've highlighted the potential of machine learning to improve 5G satellite signal decoding. We proposed two NNs to handle symbol enhancement and equalization tasks. The results demonstrate significant improvements  for both synthetic and real datasets. Our future work aims to broaden our training dataset by capturing a wider range of signals. We expect to refine our machine learning models and enhance their adaptability in real-world satellite communication environments.

\ifCLASSOPTIONcaptionsoff
  \newpage
\fi



%
%
%
\input{X.2.Acronyms}
\balance
\bibliographystyle{IEEEtran}
\bibliography{refs}

%
%
%
%
%
%




\end{document}

%% file: X.2.Acronyms.tex
\begin{acronym} [DL-OTDoA]
\acro{3G}{Third Generation}
\acro{4G}{Fourth Generation}
\acro{5G}{Fifth Generation}
\acro{6G}{Sixth Generation}
\acro{3GPP}{3rd Generation Partnership Project}
\acro{AI}{Artificial Intelligence}
\acro{AMF}{Access and Mobility Function}
\acro{AoA}{Angle of Arrival}
\acro{AR}{Augmented Reality}
\acro{AWGN}{Additive White Gaussian Noise}
\acro{BER}{Bit Error Rate}
\acro{BW}{Bandwidth}
\acro{BWP}{Bandwidth Part}
\acro{BWPP}{Bandwidth Part for Positioning}
\acro{CA}{Cell Averaging}
\acro{CDF}{Cumulative Density Function}
\acro{CIR}{Channel Impulse Response}
\acro{CFAR}{Constant False Alarm Rate}
\acro{CFO}{Carrier Frequency Offset}
\acro{CP}{Cyclix Prefix}
\acro{CP-OFDM}{Cyplix Prefix Orthogonal Frequency-Division Modulation}
\acro{CRLB}{Cramer-Rao Lower Bound}
\acro{CU}{Centralised Unit}
\acro{DDM}{Delay/Doppler Map}
\acro{DFT}{Discrete Fourier Transform}
\acro{DL-AoD}{Downlink Angle of Departure}
\acro{DL-TDoA}{Downlink Time Difference of Arrival}
\acro{DL-OTDoA}{Downlink Observed Time Difference of Arrival}
\acro{DMRS}{DeModulation Reference Signal}
\acro{DU}{Distributed Unit}
\acro{DVB}{Digital Video Broadcasting}
\acro{E-CID}{Enhanced Cell ID}
\acro{ECDF}{Empirical Cumulative Density Function}
\acro{E-SMLC}{Evolved Serving Mobile Location Center}
\acro{EIRP}{Equivalent Isotropic Radiated Power}
\acro{eNB}{evolved NodeB}
\acro{FCC}{Federal Communications Commission}
\acro{FR}{Frequency Regions}
\acro{FR1}{Frequency Region 1}
\acro{FR2}{Frequency Region 2}
\acro{GDOP}{Geometrical Dilution of Precision}
\acro{GPS}{Global Positioning System}
\acro{gNB}{Next Generation Base Station}
\acro{GNSS}{Global Navigation Satellite System}
\acro{GS}{Ground Station}
\acro{GSCN}{Global Synchronization Channel Number}
\acro{HAPS}{High-Altitude Platform Systems}
\acro{IDFT}{Inverse Discrete Fourier Transform}
\acro{IFFT}{Inverse Fast Fourier Transform}
\acro{IoT}{Internet of Things}
\acro{IIoT}{Industrial Internet of Things}
\acro{ICI}{Inter-Channel Interference}
\acro{IMU}{Inertial Measurement Unit}
\acro{ISI}{Inter-Symbol Interference }
\acro{KPI}{Key Performance Indicator}
\acro{LCS}{Location-based Services}
\acro{LEO}{Low Earth Orbit}
\acro{LMC}{Location Management Component}
\acro{LMF}{Location Management Function}
\acro{LOS}{Line of Sight}
\acro{LPP}{Localization Positioning Protocol}
\acro{LPPa}{Localization Positioning Protocol Annex}
\acro{LTE}{Long Term Evolution}
\acro{MIB}{Master Information Block}
\acro{ML}{Machine Learning}
\acro{Multi-RTT}{Multicell Round Trip Time}
\acro{NF}{Network Function}
\acro{NLOS}{Non-Line of Sight}
\acro{NN}{Neural Network}
\acro{NPRM}{Notice of Proposed Rulemaking}
\acro{NR}{New Radio}
\acro{NTN}{Non-Terrestrial Networks}
\acro{OFDM}{Orthogonal Frequency-Division Multiplexing}
\acro{OTDoA}{Observed Time Differential of Arrival}
\acro{OTFS}{Orthogonal Time Frequency Space Modulation}
\acro{PDF}{Probability Density Function}
\acro{PNT}{Positioning, Navigation, and Timing}
\acro{POD}{Precise Orbit Determination}
\acro{PPP}{Precise Point Positioning}
\acro{PRN}{Pseudo-Random Noise}
\acro{PRS}{Positioning Reference Signal}
\acro{PSS}{Primary Synchronization Signal}
\acro{PVT}{Position, Velocity, Time}
\acro{RAN}{Radio Access Network}
\acro{RAT}{Radio-Access-Technology}
\acro{RB}{Resource Block}
\acro{RE}{Resource Element}
\acro{RG}{Resource Grid}
\acro{RedCap}{Reduced Capacity}
\acro{RMSE}{Root Mean Square Error}
\acro{RTK}{Real Time Kinematics}
\acro{RRC}{Radio Resource Control}
\acro{SBAS}{Satellite Based Augmentation System}
\acro{SC}{SubCarrier}
\acro{SCS}{SubCarrier Spacing}
\acro{SIB}{System Information Block}
\acro{SIC}{Successive Interference Cancellation}
\acro{SINR}{Signal-to-Interference-plus-Noise ratio}
\acro{SLA}{Service Level Agreement}
\acro{SNR}{Signal-to-Noise Ratio}
\acro{SoO}{Signal of Opportunity}
\acro{SoP}{Signal of Opportunity}
\acro{SRS}{Sounding Reference Signal}
\acro{SSB}{Synchronization Signal Block}
\acro{SSS}{Secondary Synchronization Signal}
\acro{TA}{Timing Advance}
\acro{TDE}{Time-Domain Equalizer}
\acro{TDL}{Tapped Delay Line}
\acro{TN}{Terrestrial Network}
\acro{ToA}{Time of Arrival}
\acro{ToF}{Time of Flight}
\acro{TS}{Technical Specification}
\acro{TR}{Technical Report}
\acro{UAV}{Unmanned Aerial Vehicle}
\acro{UE}{User Equipment}
\acro{UL-AoA}{Uplink Angle of Arrival}
\acro{UL-TDoA}{Uplink Time Difference of Arrival}
\acro{VR}{Virtual Reality}
\acro{WLAN}{Wireless Local Area Network}
\acro{WLS}{Weighted Least Squares}
\end{acronym}